\documentclass[]{aastex631}
\usepackage{subfigure,graphicx} %

\usepackage{amsmath}
\usepackage{booktabs}

\def\aap{Astron.\ Astrophys.\ }

\def\apjl{Astrophys.\ J.\ Lett.\ }
\def\apjs{Astrophys.\ J.\ Supp.\ }

\newcommand{\zcr}[1]{{\color{black} {#1}}}
\newcommand{\zhu}[1]{{\color{black} {#1}}}

\begin{document}

\title {Probing  solar modulation of AMS-02 time-dependent D, $^3$He and $^4$He  fluxes with modified  force field approximation models}%


\author{Cheng-Rui Zhu}
\email{zhucr@ahnu.edu.cn}  
\affiliation{Department of Physics, Anhui Normal University, Wuhu 241000, Anhui, China }

\affiliation{Key Laboratory of Dark Matter and Space Astronomy, Purple
Mountain Observatory, Chinese Academy of Sciences, Nanjing 210008, Jiangsu, China }

\author{Mei-Juan Wang}
\affiliation{Department of Physics, Anhui Normal University, Wuhu 241000, Anhui, China }

\date{\today}

\begin{abstract}

The AMS-02 experiment recently published time-dependent fluxes of deuterons (D) from May 2011 to April 2021, divided into 33 periods of four Bartels rotations each. These temporal structures are associated with solar modulation.   In this study, three modified force-field approximation are employed to examine the long-term behavior of cosmic-ray (CR) isotopes such as D, $^3$He, and $^4$He, as well as the ratios D/$^3$He and $^3$He/$^4$He. The solar modulation potential is rigidity-dependent for these modified force-field approximation models.  Due to the unknown local interstellar spectrum (LIS) for these isotopes, we utilize the Non-LIS method for solar modulation. By fitting to the AMS-02 time-dependent fluxes, we derive the solar modulation parameters. Our findings prove the assumption in literature that all isotopes can be fitted using the same solar modulation parameters and it shown that the modified FFA models are validated parametrization for solar modulation. Based on these, we forecast the daily fluxes of D, $^3$He and $^4$He from 2011 to 2020.



\end{abstract}



\section{Introduction}

Galactic cosmic rays (GCRs) originate from cosmic accelerators such as the shocks of supernova remnants and then propagate within the Milky Way \citep{1998ApJ...493..694M}. However, when they traverse the heliosphere to Earth, they are modulated by the outward-moving magnetized solar wind plasma \citep{Potgieter2013}.
Solar modulation is crucial for understanding the nature of GCRs, including their origin \citep{2013A&ARv..21...70B} and propagation \citep{2007ARNPS..57..285S} within the galaxy. It is also significant for searching for dark matter through low-energy modulated antiproton and antideuteron fluxes \citep{2012CRPhy..13..740L,Yuan:2014pka,2017PhRvL.118s1101C,PhysRevLett.129.091802,2022PhRvL.129w1101Z}.
Previous studies have shown that solar activity influences solar modulation \citep{2011JGRA..116.2104U,Potgieter2013}. Stronger solar activities lead to higher levels of solar modulation. The solar modulation exhibits periodicities of 11-year and 22-year cycles corresponding to solar activity periods \citep{AMS:2021qln,PhysRevLett.128.231102} and daily fluxes show a 27-day cycle corresponding to the solar rotation \citep{AMS:2021qln,PhysRevLett.128.231102}.

To model and predict the intensity of Galactic Cosmic Rays (GCRs), the Parker's transport equation (TPE) \citep{1965P&SS...13....9P} is employed to depict the propagation processes of cosmic rays within the heliosphere. The Parker’s transport equation can be solved by numerical methods or analytical methods. Usually, the force field approximation (FFA) \citep{1967ApJ...149L.115G,1968ApJ...154.1011G} is used to solve the equation as it is simple and enough to explain the observations. 
However, with the development of instruments, such as PAMELA, AMS-02, and DAMPE \citep{2011Sci...332...69A,2017PhRvL.119y1101A,cite-key}, the observation has entered a high-precision era.
Alhough the use of FFA is usually employed for
most models of Galactic cosmic rays, it is inadequate to describe the time-dependent GCR spectra themselves as well as the fluxes ratio. 
\cite{PhysRevLett.121.251104,Corti_2019,Song_2021,2022PhRvL.129w1101Z,2022PhRvD.106f3006W} reproduced the AMS-02 observations using a one-dimensional or a three-dimensional numerical model respectively to solve the Parker equation. Several methods have been proposed to modify the  FFA (\citep{2016ApJ...829....8C,2017PhRvD..95h3007Y,Zhu:2020koq,2021ApJ...921..109S,PhysRevD.109.083009,Li_2022}) to account for the cosmic rays fluxes.

The recent experimental results have achieved significant breakthroughs that are instrumental in understanding the solar modulation effect as well as the GCRs physics, particularly those from Voyager and AMS-02. So far, only Voyager-1 and Voyager-2 have crossed the boundary of the heliosphere \citep{2013Sci...341..150S,voyager-2} and detected the LIS in the range from a few to hundreds MeV/nucleon. Unfortunately, it is challenging for them to distinguish particle isotopes, which makes it difficult for us to get the LIS of non-dominant isotopes. The AMS-02 collaboration has published a variety of high-precision cosmic ray spectra \citep{AGUILAR20211} and the evolution of some cosmic ray fluxes over time \citep{PhysRevLett.121.051101,PhysRevLett.121.051102}. 

Very recently, the AMS-02 experiment released the time-dependent fluxes of D, $^3$He, and $^4$He \citep{PhysRevLett.132.261001} from May 2011 to April 2021, providing new opportunities to study the solar modulation for cosmic rays (CRs). In this study, we employ three modified force field approximation models from previous research to investigate the solar modulation of D, $^3$He, and $^4$He. 
\zcr{It is assumed that each of these modified force-field approximation models incorporates a rigidity-dependent solar modulation potential.}
In order to study the solar modulation of Galactic Cosmic Rays (GCRs), it is necessary to assume Local Interstellar Spectra (LIS). Typically, Voyager data are used to constrain the LIS. However, the Voyager  does not provide data for D, $^3$He, and $^4$He.  To eliminate the impact of LIS spectra, an alternative method was proposed in \citep{corti2019testvalidityforcefieldapproximation}, referred to here as the Non-LIS method. The results of this study confirm that the modified FFA model represents a valid parameterization of the spectrum for various GCRs.
This work is the first in the literature to show that other nuclei species can share the same solar modulation parameters with protons and helium to interpret the long-term behavior of GCR fluxes.
It is of great utility in elucidating the long-term behavior of GCRs and predicting the long-time fluxes of unmeasured GCRs. Based on this assumption, we predict the daily fluxes of D, $^3$He and $^4$He from 2011 to 2020, and most of  the total fluxes of $^3$He and $^4$He agree with  the measurements of He within their 1 $\sigma$  confidence interval.

\section{Methodology}
\subsection{Solar Modulation}

The existence of heliospheric magnetic field carried by solar winds causes the modulation of GCRs as they enter the heliosphere, resulting in suppressed fluxes of CRs. 
Above about 30 GeV/n,  this effect is negligible but it gets increasingly  more pronounced at lower energies \citep{Potgieter2013}.
The basic transport equation (TPE) was first derived by Parker \citep{1965P&SS...13....9P,Potgieter2013} as shown in Eq. \ref{parker}
\begin{equation}\label{parker}
\begin{aligned}
\frac{\partial f}{\partial t} = &-(\vec V_{sw}+<\vec v_D>) \cdot \nabla  f + \nabla \cdot (\bold K^{(s)} \cdot \nabla f )\\
&+ \frac{1}{3}(\nabla\cdot \vec V_{sw})\frac{\partial f}{\partial ln p}.
\end{aligned}
\end{equation}
Here, $f$ is the cosmic ray distribution function in the phase space $(\vec r,p)$, and $p$ denotes momentum. The cosmic ray flux $J$ measured by
experiments is related to the distribution function by$ J \propto p^2 f$.  Particle rigidity
$R$ which is widely used in experimental studies, is related to momentum by $R = pc/q$, where $c$ is the speed of light and $q$ is
the charge of the cosmic ray particles. $\vec V_{sm}$ is the solar wind speed, $\vec v_d$ is the drift velocity and $\bold K^{(s)}$ denotes the diffusion tensor. The terms in the right of Eq. \ref{parker} describe the convection, drift, diffusion and adiabatic energy loss of GCRs transport effects in the heliosphere. 
There are various methods to solve the equation and  the force-field approximation (FFA) \citep{1967ApJ...149L.115G, 1968ApJ...154.1011G} is mostly applied as it is simple enough.
In this model,  it is assumed that spherical symmetry with (a) a steady state ($\partial f /\partial t = 0$), (b) an adiabatic energy loss rate ⟨$<dP/dt> = (P/3)\vec V_{sw} \cdot \nabla f / f = 0$, and (c) no drifts.  and we get  that the TOA flux is related with the LIS flux as 
\begin{equation}\label{force_filed}
J^{\rm TOA}(E)=J^{\rm LIS}(E+\Phi)\times\frac{E(E+2m_p)}
{(E+\Phi)(E+\Phi+2m_p)}, 
\end{equation}
where $E$ is the kinetic energy per nucleon, $\Phi=\phi\cdot Ze/A$ with 
$\phi$ being the solar modulation potential, $m_p=0.938$ GeV is the 
proton mass, and $J$ is the differential flux of GCRs. The only
parameter in the force-field model is the modulation potential $\phi$.


\subsection{The Non-LIS method and modified FFA models}

Usually, we need the LIS  to study the time-dependent solar modulation effects.  However we know very very little about the LIS of D, $^4$He and $^3$He as  there have been very few experiments to detect these cosmic rays.  In order to eliminate the influence of CR LIS, the Non-LIS method is adopted here. If we have two modulated spectral $J^{TOA}(t_1)$ and $J^{TOA}(t_2)$, we get the relation of them from Eq. \ref{force_filed} as \citep{corti2019testvalidityforcefieldapproximation}
\begin{equation}\label{force_filed2}
\begin{aligned}
J^{\rm TOA}(E,t_1)=&J^{\rm TOA}(E+\Delta \Phi,t_2) \\
&\times \frac{E(E+2m_p)}{(E+\Delta\Phi)(E+\Delta\Phi+2m_p)}.
\end{aligned}
\end{equation}
Here,  $\Delta\Phi =\Phi(t_1) - \Phi(t_2) = Ze/A \cdot (\phi(t_1) - \phi(t_2)) = 
Ze/A \cdot \Delta \phi$.  In this work,  we take the mean fluxes from  May 2011 to April 2021 as the $J^{TOA}(t_2)$ \citep{PhysRevLett.132.261001}. 
We utilized linear interpolation on a logarithmic scale (log R - log J) to determine the interpolated flux values for other rigidities that lack observational data.

In principle, the force-field model assumes a quasi-steady-state of 
the solution of the Parker's equation. However, the observational GCRs 
fluxes show 11-year variations associated with solar activities.
Therefore a time-series of $\phi$ at different epochs is adopted to
describe the data. As with only one parameter, we can not fit the time-dependent cosmic rays fluxes very well, rigidity-dependent solar modulation potential is needed \citep{SIRUK20241978}. In this study, we adopt three modified force-field models. First, we employ the modified force-field model (Zhu's model) from \cite{zhu2024}, which is an extension of the model presented in \cite{2016ApJ...829....8C,2017JGRA..12210964G}. The solar modulation potential is
\begin{equation}\label{Zhu}
\Delta \phi (R)_{Zhu} = \phi_l +\left (\frac{\phi_h-\phi_l}{1+e^{(-R+R_b)}} \right ),
\end{equation}
where $\phi_l$ is the solar modulation potential for the low energy, and $\phi_h$ is for the high energy, $e$ is the natural constant, $R$ is the rigidity and $R_b$ is the break rigidity. The sigmoid function is employed here to smooth the transition. $\phi_l$, $\phi_h$ and $R_b$ are the free parameters to be fitted.
Second, we take the Cholis’ model from \cite{2016PhRvD..93d3016C,Cholis:2020tpi} as
\begin{equation}\label{Cholis}
\Delta \phi (R)_{Cholis} = \phi_0 + \phi_1 \left(  \frac{1+(R/R_0)^2}{\beta(R/R_0)^3}   \right) .
\end{equation}
Here, $\beta$ is the  ratio between the particle speed and the speed of light. $\phi_1$, $\phi_2$ and $R_0$ are the free parameters to be fitted.
Third, we take  Long's model from \cite{PhysRevD.109.083009} as 
\begin{equation}\label{Long}
\Delta \phi (R)_{Long} = \phi_0 + \phi_1 ln(R/R_0), 
\end{equation}
with 
\begin{equation}\label{force_filed3}
\begin{aligned}
J^{\rm TOA}(E,t_1)=&J^{\rm TOA}(E+\Delta \Phi,t_2) \\
&\times \frac{E(E+2m_p)}{(E+\Delta\Phi)(E+\Delta\Phi+2m_p)}\\
&exp(-g\frac{10R^2}{1+10R^2}\Delta\phi)).
\end{aligned}
\end{equation}
Here, $\phi_0$, $\phi_1$ and $g$ are the free parameters to be fitted.  \cite{2021ApJ...921..109S} presents Shen's model. However, to achieve better fitting results for different cosmic-ray species, it is necessary to modify the model to be rigidity dependent rather than energy dependent. Therefore, we do not employ their model in this work.

\subsection{MCMC}

We fit the three solar modulation $\phi(R)$ with three free parameters. The $\chi^2$ statistics is defined as
\begin{eqnarray}
\chi^2=\sum_{i=1}^{m}\frac{{\left[J(E_i;\phi(R))-
J_i(E_i)\right]}^2}{{\sigma_i}^2},
\end{eqnarray}
where $J(E_i;\phi(R))$ is the expected modulated flux, $J_i(E_i)$ and
$\sigma_i$ are the measured flux and error for the $i$th data bin with the geometric mean of the bin edges as $E_i$.

We use the Markov Chain Monte Carlo (MCMC) algorithm to  minimize the $\chi^2$ function, which works in the Bayesian framework.  The posterior
probability of model parameters $\boldsymbol{\theta}$ is given by
\begin{equation}
p(\boldsymbol{\theta}|{\rm data}) \propto {\mathcal L}(\boldsymbol{\theta})
p(\boldsymbol{\theta}),
\end{equation}
where ${\mathcal L}(\boldsymbol{\theta})$ is the likelihood function
of parameters $\boldsymbol{\theta}$ given the observational data, and 
$p(\boldsymbol{\theta})$ is the prior probability of $\boldsymbol{\theta}$.

The MCMC driver is adapted from {\tt CosmoMC} \citep{2002PhRvD..66j3511L,Liu_2012}.
We adopt the Metropolis-Hastings algorithm. The basic procedure of this
algorithm is as follows. We start with a random initial point in the 
parameter space, and jump to a new one following the covariance of these
parameters. The accept probability of this new point is defined as
$\min\left[p(\boldsymbol{\theta}_{\rm new}|{\rm data})/p(\boldsymbol{\theta}_
{\rm old}|{\rm data}),1\right]$. If the new point is accepted, then repeat
this procedure from this new one. Otherwise go back to the old point.
For more details about the MCMC one can refer to \citep{MCMC}.

\section{results and  discussion }

\begin{figure*}
    \centering
    \includegraphics[scale=0.7]{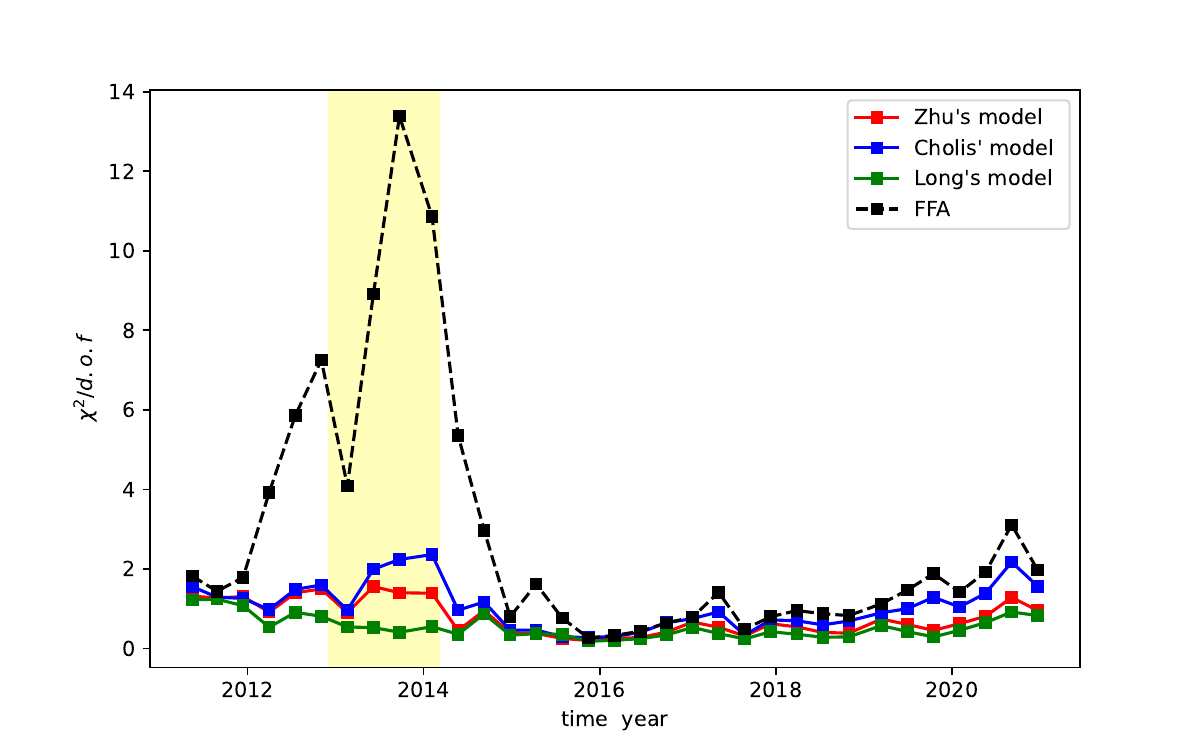}
    \caption{The figure illustrates $\chi^2/d.o.f$ values for the four models for the  combined analysis (D+$^3$He+ $^4$He). Specifically, the blue curve corresponds to Zhu's model, the red curve represents Cholis' model, and the green curve denotes Long's model. The  FFA result is depicted in black for comparison.
    The yellow shaded band stands for the   heliospheric magnetic field reversal period within which the polarity is uncertain\citep{2015ApJ...798..114S}.}
    \label{fig:chi2}
\end{figure*}

\begin{figure*}
    \centering
    \includegraphics[scale=0.6]{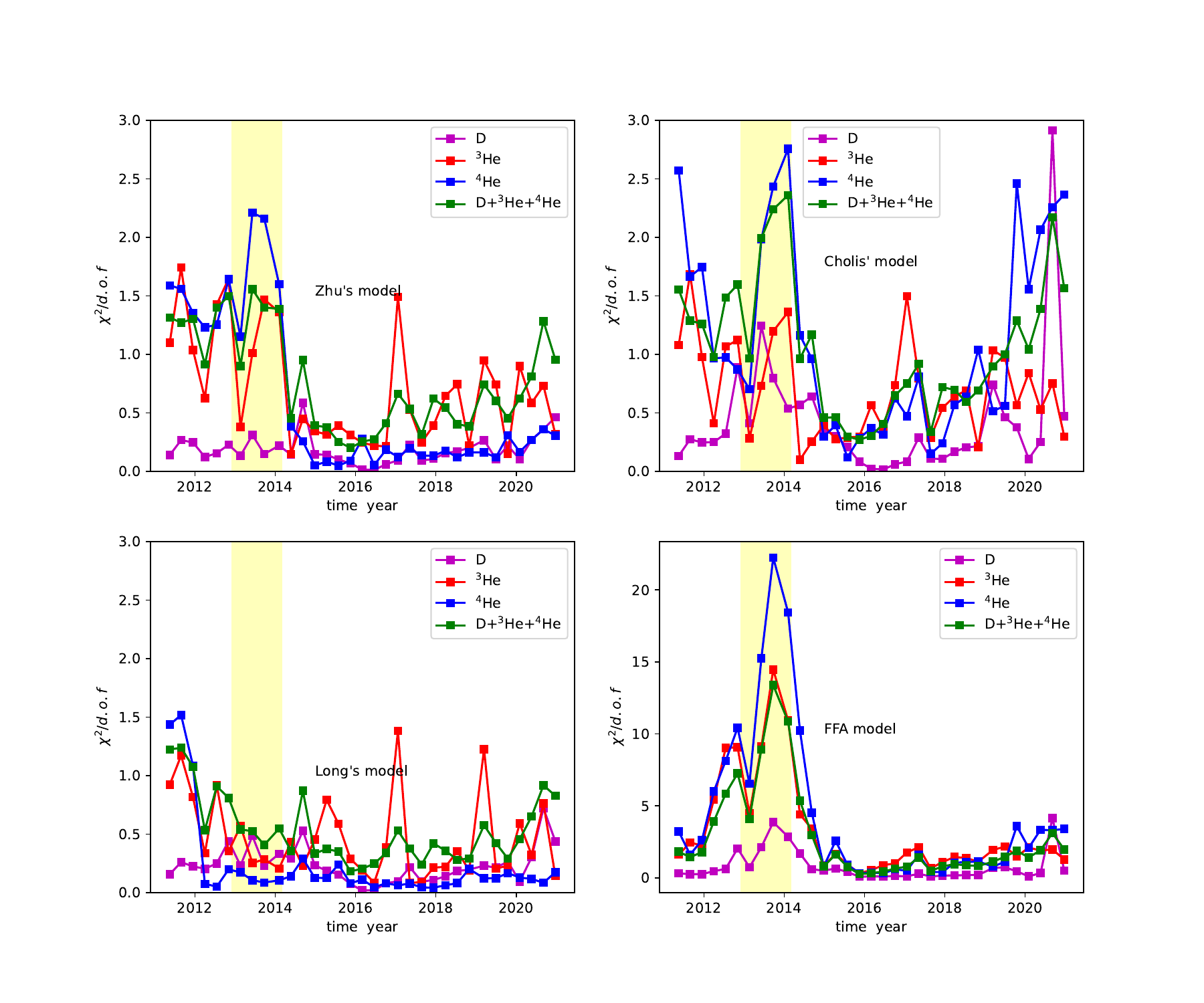}
    \caption{The $\chi^2/d.o.f$ over time  for the analysis of each independent isotope (D, $^3$He or $^4$He), as well as the combined
analysis (D+$^3$He+ $^4$He),    for the Zhu's model (top left), the Cholis' model (top right), the Long's model (bottom left) and the FFA model (bottom right). In the FFA model, the degrees of freedom ($d.o.f$) are 25 for the analysis of each independent isotope and 77 for the combined analysis. As for the other three models, the $d.o.f$ are 23 for the analysis of each independent isotope and 75 for the combined analysis.
    The yellow shaded band stands for the   heliospheric magnetic field reversal period within which the polarity is uncertain\citep{2015ApJ...798..114S}.}
    \label{fig:chi2_all}
\end{figure*}

\begin{figure*}
    \centering
    \includegraphics[width=1.0\linewidth]{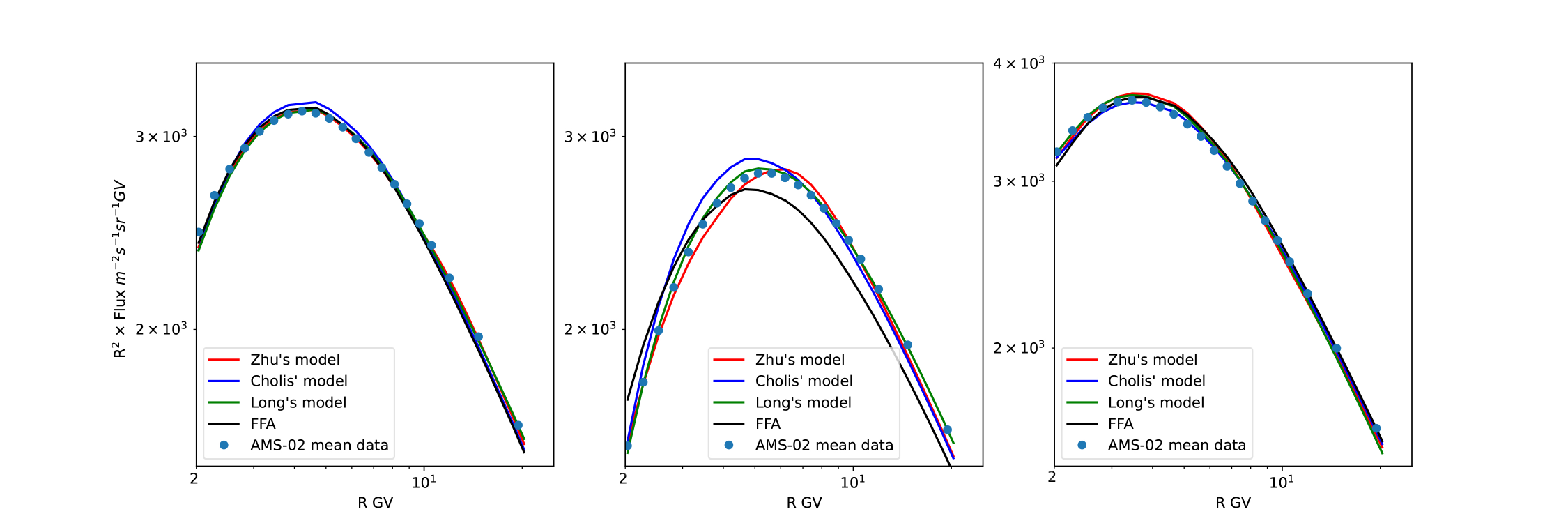}
    \caption{The model prediction proton fluxes from 2011-05-20 to 2011-08-30, from 2014-02-07 to 2014-05-2 and from 2017-05-10 to 2017-08-25 respectively with the same parameters from the fitting in the manuscript.  The $J^{TOA}(t_2)$ in the Eq.\ref{force_filed2} here for proton is derived from the $\Phi_D$/(ratio of D to p) from \citep{PhysRevLett.132.261001}. The point data is taken from the mean fluxes of AMS-02 daily protons fluxes  \citep{AMS:2021qln} in the same period.   \zcr{Note that the Y-axis represents the flux, which we have re-scaled using the $R^2$ \citep{1995NIMPA.355..541L}} }
    \label{fig:p}
\end{figure*}

\begin{figure*}
    \centering
    \includegraphics[scale=0.7]{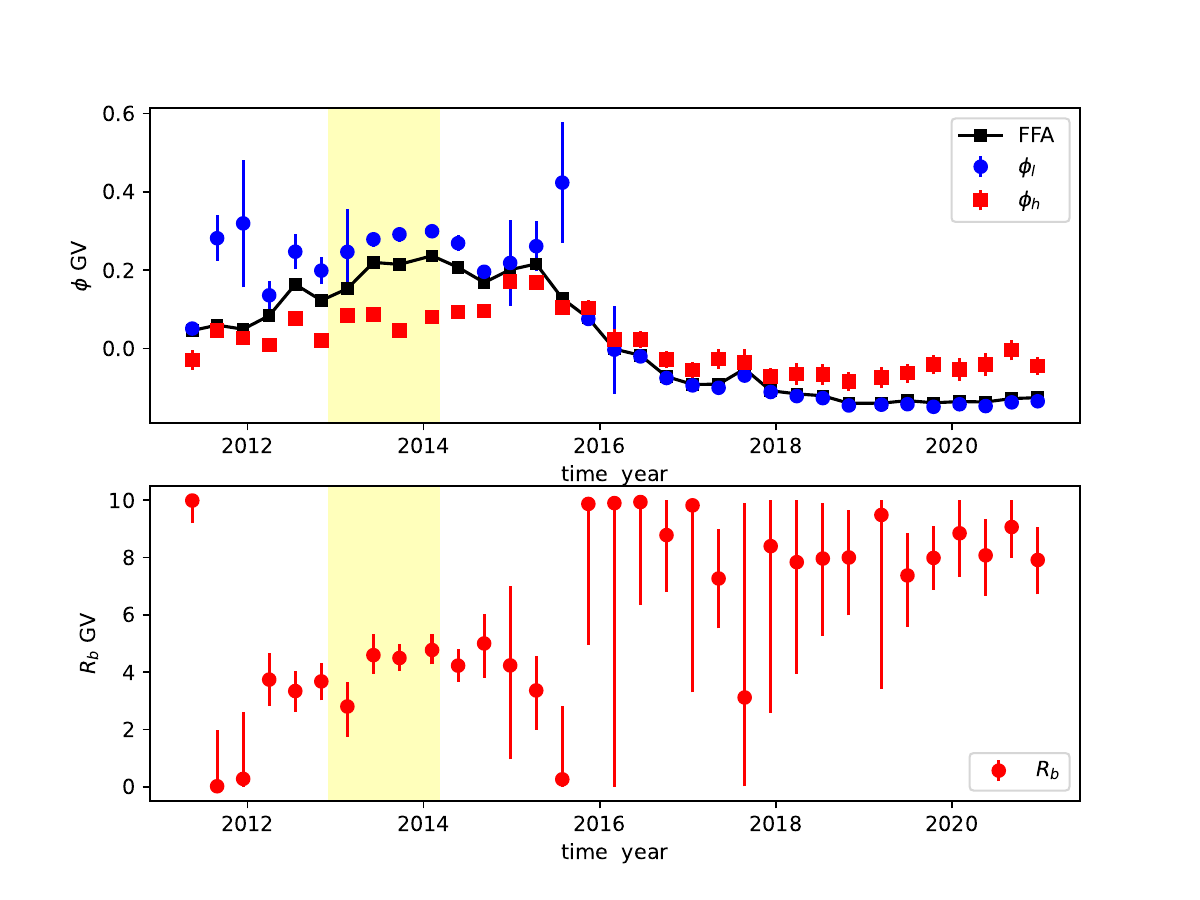}
    \caption{The fitting results of Zhu's model.  (Top) Time series of $\phi_l$ (blue)  and $\phi_h$ (red) via fitting to the AMS-02 data. \zcr{The solar modulation potentials from the FFA (black) are shown for comparison.} (Bottom) Same to the top but for $R_b$.  
    The shaded band stands for the   heliospheric magnetic field reversal period within which the polarity is uncertain.}
    \label{fig:phi_zhu}
\end{figure*}

We have jointly fitted the time-dependent spectra of D, $^3$He, and $^4$He from AMS-02 together with FFA and three modified FFA models. The time dependent $\chi^2/d.o.f$ values are shown in Fig. \ref{fig:chi2}. It is noteworthy that the degrees of freedom (d.o.f) for the FFA are 77, whereas for the three modified FFA models, they amount to 75. It is evident that the three modified FFA  models significantly reduce the $\chi^2/d.o.f$ during the period of heliospheric magnetic field reversal, when the polarity remains uncertain. Meanwhile, the  FFA continues to provide adequate predictions during the solar minimum from 2015 to 2021. Upon comparing the three modified FFA models, we find that Long's model yields the best fitting results, with $\chi^2/d.o.f$ values ranging from 0.182 to 1.239 and a mean value of 0.537. Zhu's model provides $\chi^2/d.o.f$ values ranging from 0.202 to 1.557 and a mean value of 0.771. Meanwhile, Cholis' model gives $\chi^2/d.o.f$ values ranging from 0.270 to 2.361 and a mean value of 1.054. 

In Fig. \ref{fig:chi2_all}, we show the $\chi^2$ results of the  four models for independent analysis  of D, $^3$He and $^4$He, as well as the combined analysis respectively. The specific values are shown in Table. \ref{tab:zhu}, \ref{tab:Cholis}, \ref{tab:Long} and \ref{tab:FFA} in the appendix. Similar to the combine analysis, the FFA is particularly poor to simulate the solar modulation during  the heliosphere magnetic field reversal period for the independent analysis. So that we need a more reliable model to study the propagation of Galactic cosmic rays. As the relative error of the D flux is 2 to 3 times that of $^3$He and $^4$He, the deuterium gets  better fitting results than He isotopes. It means  that for more precise data, we may need a more refined and physically motivated model. However, for now, the Zhu and Long's  models are sufficient because the average $\chi^2/d.o.f$ value is smaller than 1.

\zcr{In Fig. \ref{fig:p}, we show the model prediction proton fluxes from 2011-05-20 to 2011-08-30, from 2014-02-07 to 2014-05-2 and from 2017-05-10 to 2017-08-25 respectively with the same parameters from the fitting in the manuscript with four models.} The $J^{TOA}(t_2)$ in the Eq.\ref{force_filed2} for proton here is derived from the $\Phi_D$/(ratio of D to p) from \citep{PhysRevLett.132.261001}. \zcr{First, we got the $\Phi_p$ for each time interval reported in \cite{PhysRevLett.132.261001}, then  derived the average value of these $\Phi_p$, which we designate as $J^{TOA}(t_2)$.} \zcr{The point data is taken from the mean fluxes of AMS-02 daily protons fluxes  \citep{AMS:2021qln} in the same period. Note that the proton publication of \cite{AMS:2021qln} includes the D flux, i.e. p+D, in this paper the D component is not excluded.} During the heliospheric magnetic field reversal period, within which the polarity is uncertain, Long's model gives the best prediction results, followed by Zhu's and Cholis' models. The FFA result is far away from the measurement. For the three modified models, the predicted proton fluxes correspond to the measurements within about 5\% for most cases.

All  the three modified models incorporate a rigidity-dependent solar modulation potential and yield very good fitting results. Although the Long's model gets the best fitting results, it  may not be  suitable to extend to high rigidity. The scale index g potentially exerts a consistent and uniform influence across all rigidity ranges, essentially resembling a rescaling of the LIS, where it efficiently allocates distinct LIS values to varying epochs. In this work, we find that the Cholis' model is only slightly worse than Zhu's model. However, \citep{PhysRevD.109.083009} shows that Cholis' model performs poorly in simulating the solar modulation for proton and helium fluxes, particularly during the solar reversal phase. As the rigidity range of deuterium,$^3$ He, and $^4$He is 2 to 20 GV, while the rigidity range for proton and helium is 1 to 60 GV, this may indicate that the Cholis' model is only suitable for a narrower rigidity range.

 The main reason  of rigidity-dependence of solar modulation potential is the break in the  diffusion coefficients.  The typical empirical expression of diffusion coefficients used in numerical are broken power-law, where $a$ for the slope of the power law at low rigidities  and $b$ for the slope of the power law at high rigidities \citep{2014SoPh..289..391P}.  \citep{2014SoPh..289..391P,2015ApJ...810..141P,2017ApJ...834...89D,Luo_2019,Song_2021} modeled the time-dependent GCRs spectra by adjusting the diffusion coefficients with time. The different slope of the power law at low and high rigidities will induce a rigidity-dependent solar modulation behavior and the modified FFA models here can empirically describe this behavior.


The fitting parameters for Zhu's model are shown in Fig. \ref{fig:phi_zhu}.
In the bottom panel of Fig. \ref{fig:phi_zhu}, we present the time-series of $R_b$. The mean value of $R_b$ is 6.03 GV, corresponding to the result of \cite{zhu2024} within 2 $\sigma$ confidence intervals. Notably, if the value of $R_b$ is excessively low, it implies that the potential transition range will be more significant in the low energy region, rather than being dependent on $\phi_l$. The time-series of $\phi_l$ and $\phi_h$ are displayed in the top panel of Fig. \ref{fig:phi_zhu}. The fluctuation of $\phi_h$ is relatively small, with a mean value of 0.014 GV, ranging from $-$0.085 GV to 0.174 GV. This is because high energy particles are less influenced by solar modulation. 
 
Around 2016, the value of $\phi_l$  approaches $\phi_h$,   indicating that the  FFA is sufficient to explain solar modulation and the break is not obvious, so that the parameter $R_b$ is not important. The model indeed shows signs of overfitting here.  These overfitting means that we only need one parameter to describe the solar modulation here, and the solar modulation potential seems to be rigidity-independent. But we should notice that we use the none LIS solar modulation model here, which means the solar modulation rigidity-dependence around 2016 should be same or similar to the solar modulation potential of $J^{TOA}(t_2) $ used in the paper. 
  
\begin{figure}
    \centering
    \includegraphics[scale=0.6]{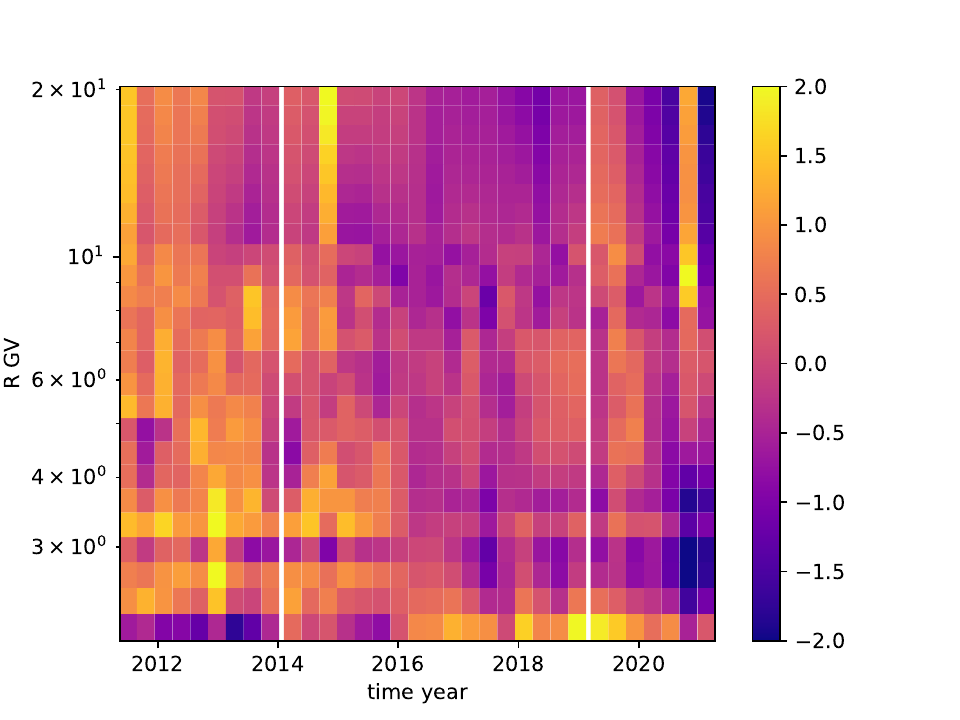}
    \caption{Zhu's model prediction comparing to the data ($\frac{J_{model}-J_{data}}{\sigma_{data}}$)  of D from May 2011 to April 2021.}
    \label{fig:D}
\end{figure}

\begin{figure}
    \centering
    \includegraphics[scale=0.6]{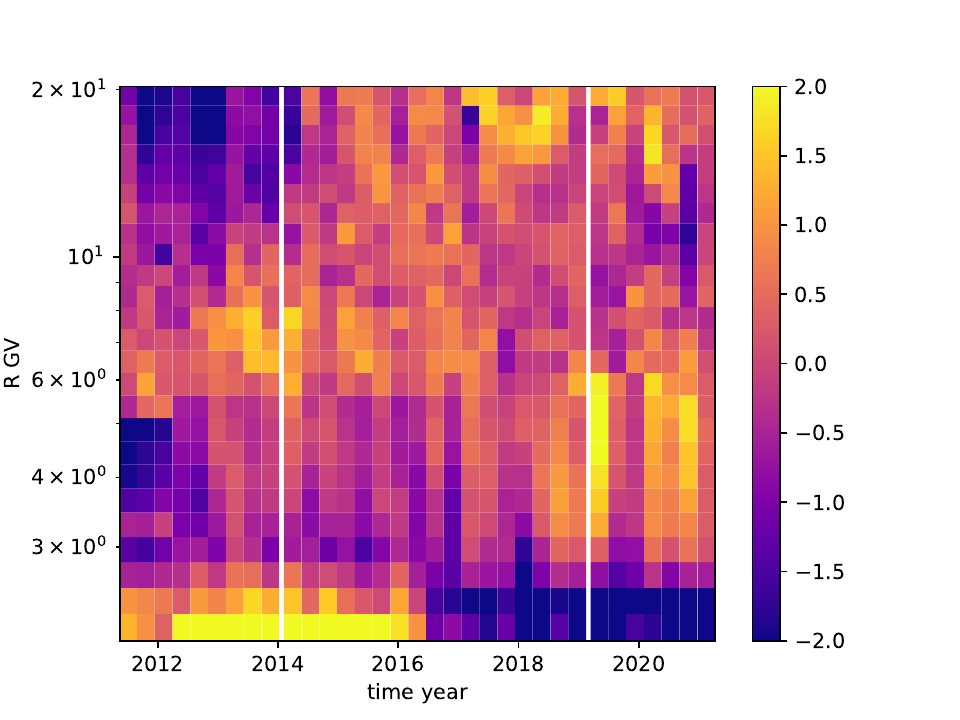}
    \caption{Zhu's model prediction comparing to the data  ($\frac{J_{model}-J_{data}}{\sigma_{data}}$)
 of $^3$He from May 2011 to April 2021.}
    \label{fig:He3}
\end{figure}

\begin{figure}
    \centering
    \includegraphics[scale=0.6]{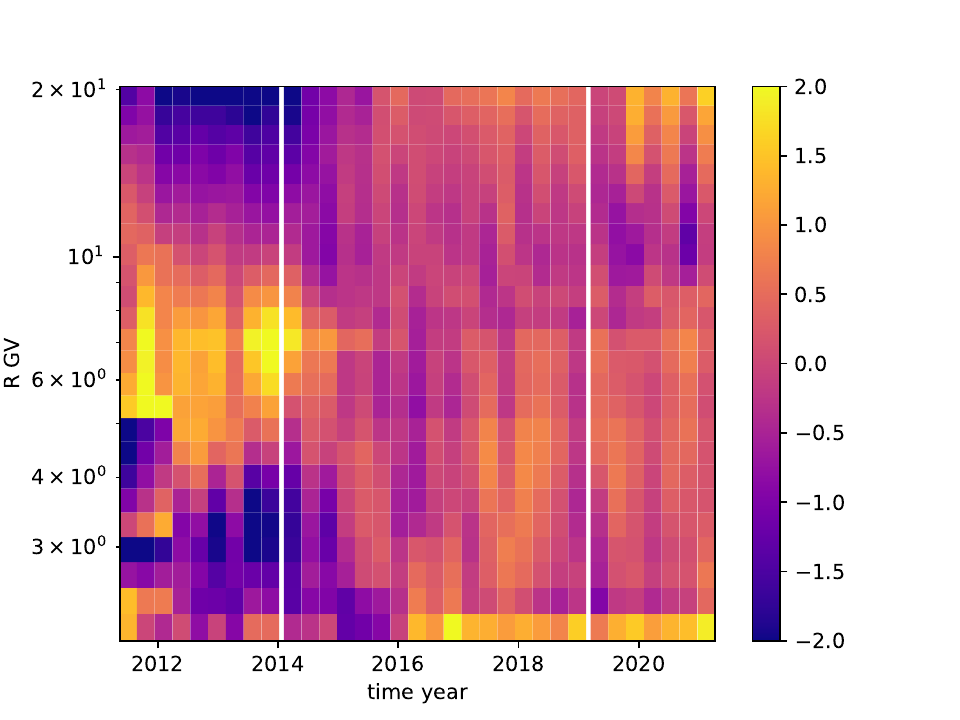}
    \caption{Zhu's model prediction comparing to the data ($\frac{J_{model}-J_{data}}{\sigma_{data}}$) of $^4$He from May 2011 to April 2021.}
    \label{fig:He4}
\end{figure}

In Fig. \ref{fig:D}, \ref{fig:He3}, \ref{fig:He4}, we present the ratios of Zhu's model-predicted intensities to measured values for D, $^3$He and $^4$He , respectively.  
It is evident that most of the fits align with the data within a 2$\sigma_{data}$ margin, utilizing the same solar modulation parameters. 

\begin{figure*}
    \centering
    \includegraphics[scale=0.6]{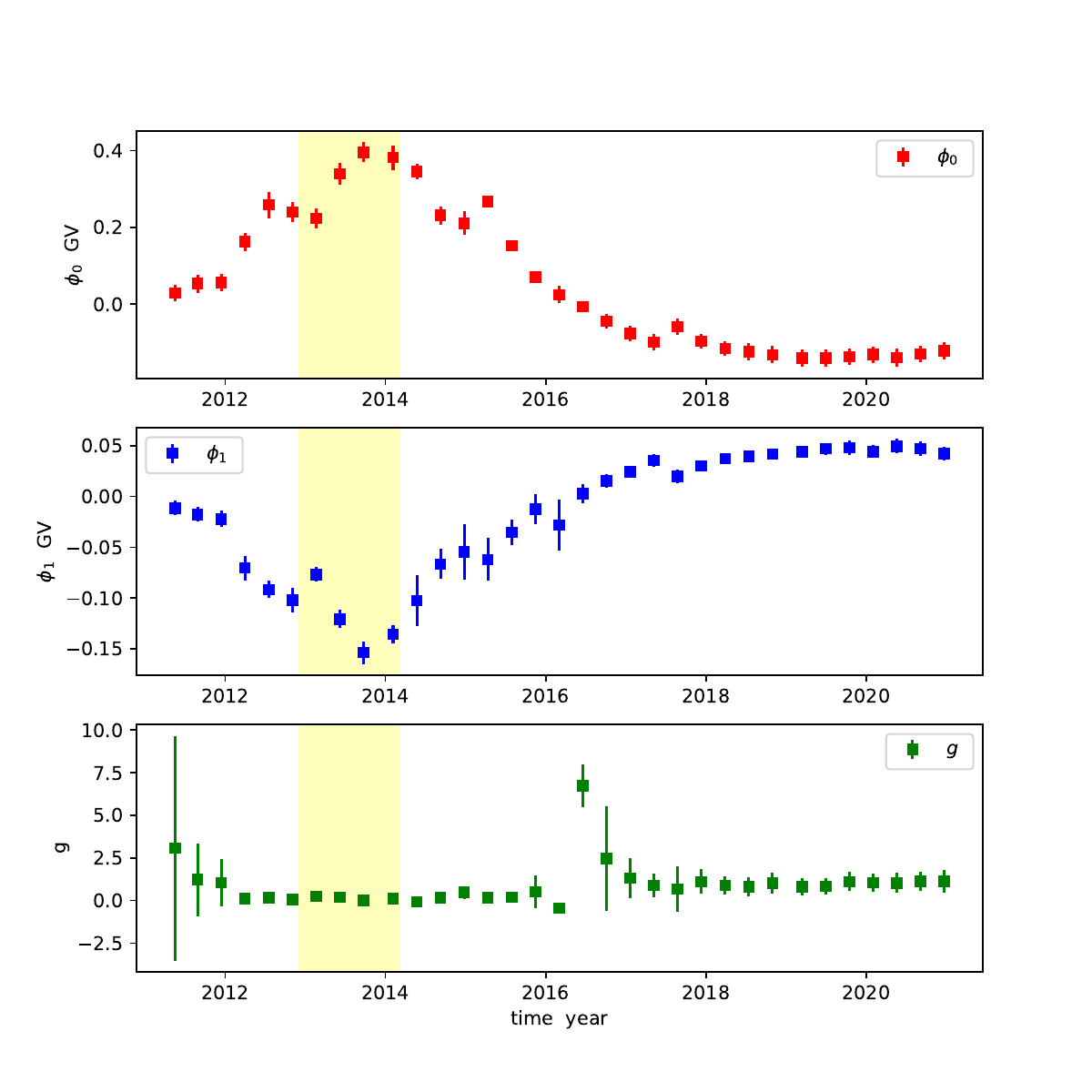}
    \caption{The fitting results of Long's model.  (Top) Time series of $\phi_0$.   (Middle) Time series of $\phi_1$. (Bottom) Same to the top but for $g$. 
    The shaded band stands for the   heliospheric magnetic field reversal period within which the polarity is uncertain.}
    \label{fig:phi_long}
\end{figure*}

\begin{figure}
    \centering
    \includegraphics[scale=0.6]{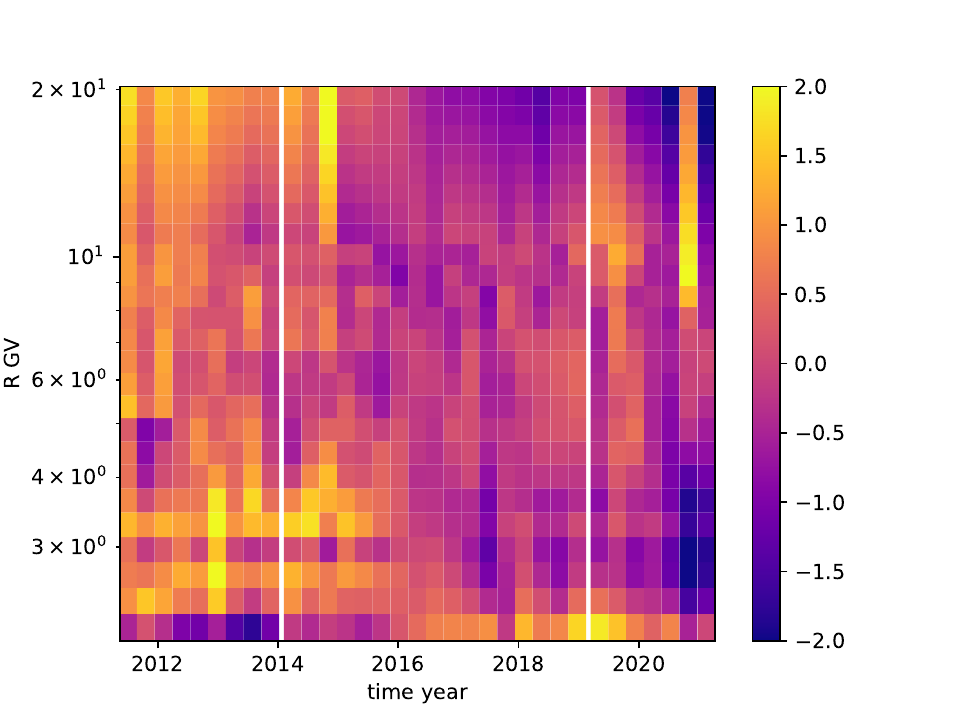}
    \caption{Long's model prediction comparing to the data ($\frac{J_{model}-J_{data}}{\sigma_{data}}$)  of D from May 2011 to April 2021.}
    \label{fig:Dl}
\end{figure}

\begin{figure}
    \centering
    \includegraphics[scale=0.6]{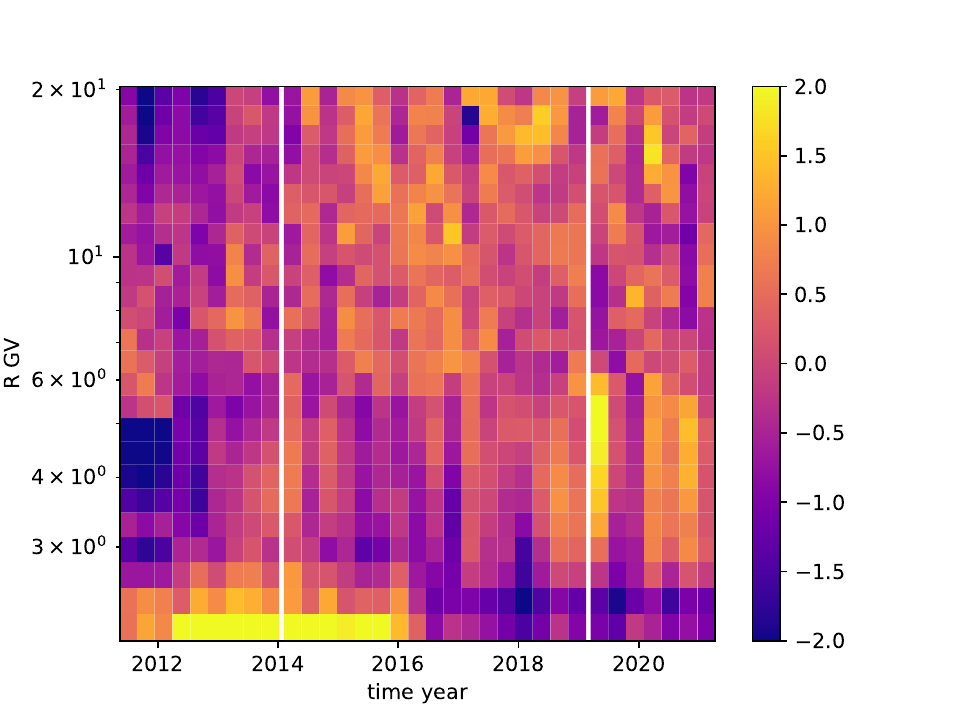}
    \caption{Long's model prediction comparing to the data  ($\frac{J_{model}-J_{data}}{\sigma_{data}}$)
 of $^3$He from May 2011 to April 2021.}
    \label{fig:He3l}
\end{figure}

\begin{figure}
    \centering
    \includegraphics[scale=0.6]{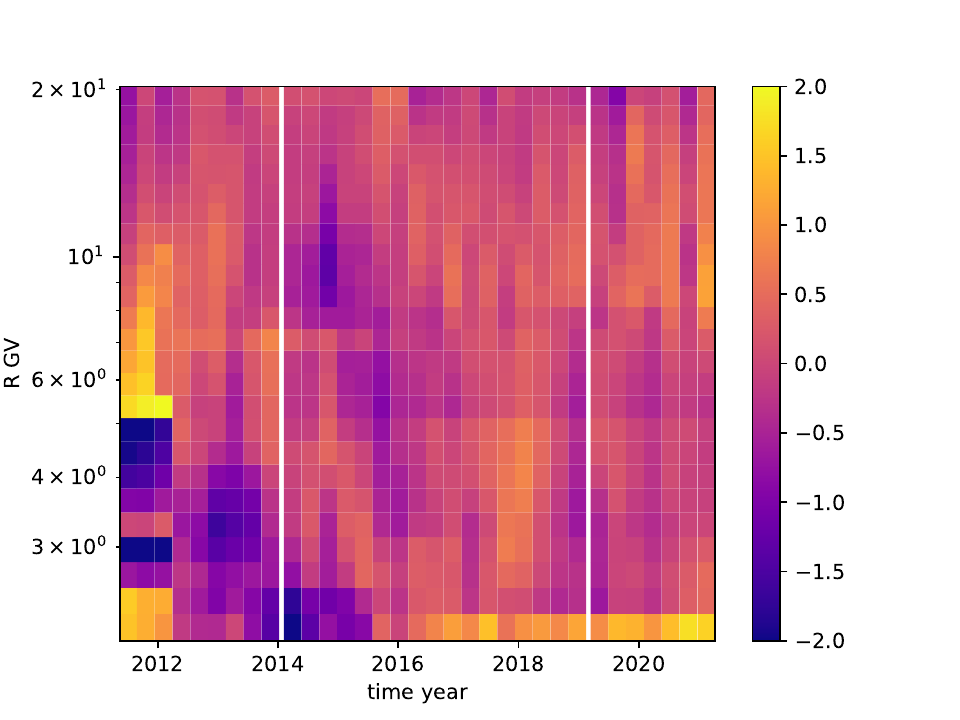}
    \caption{Long's model prediction comparing to the data ($\frac{J_{model}-J_{data}}{\sigma_{data}}$) of $^4$He from May 2011 to April 2021.}
    \label{fig:He4l}
\end{figure}

We show the fitting parameters for Long's model in Fig. \ref{fig:phi_long}. Before 2016, $\phi_1$ is less  than zero, and after 2016, $\phi_1$ is greater than zero. So the rigidity-dependence is similar to the Zhu's model. In Fig. \ref{fig:Dl}, \ref{fig:He3l}, and \ref{fig:He4l}, we show the Long's model predictions compared to measurements.  Here we can see that Long's model is slightly better than Zhu's model.

\begin{figure*}
    \centering
    \includegraphics[scale=0.6]{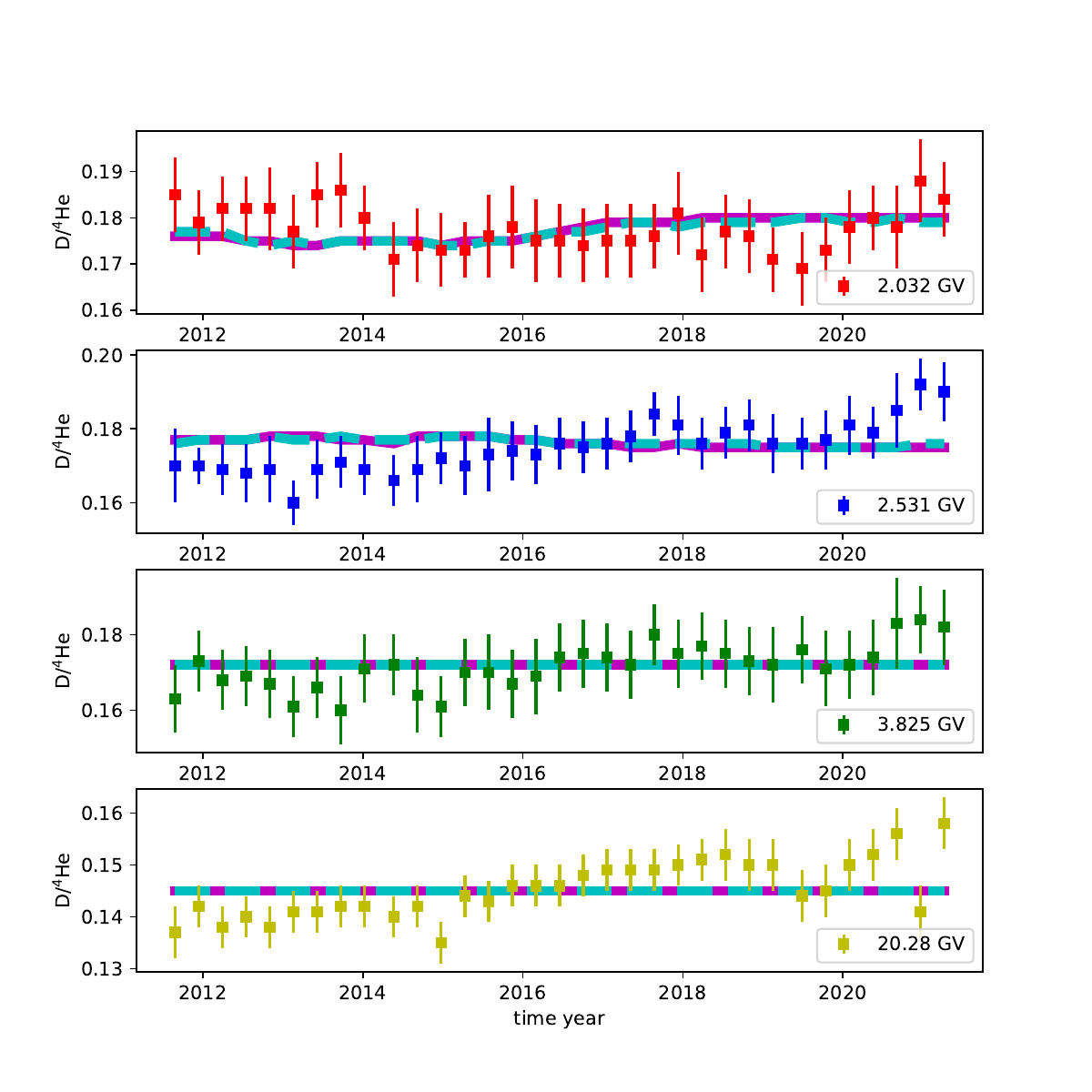}
    \caption{ Zhu's model prediction (magenta line) and Long's model prediction (cyan line) of D/$^4$He fluxes ratio comparing to the data from May 2011 to April 2021 at rigidities = 2.032 GV, 2.531 GV, 3.825 GV and 20.28 GV.}
    \label{fig:DHe4}
\end{figure*}

\begin{figure*}
    \centering
    \includegraphics[scale=0.6]{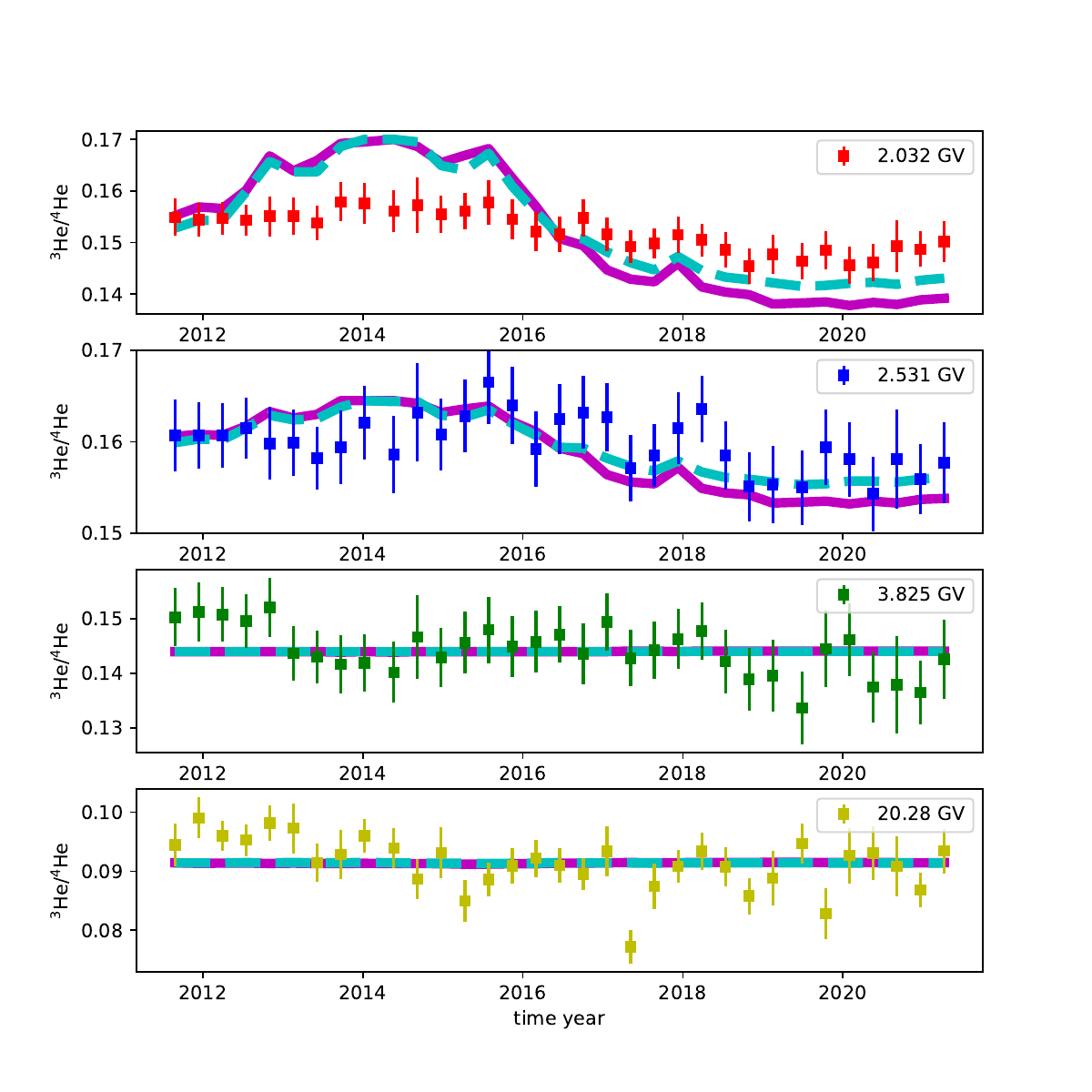}
    \caption{Zhu's model prediction (magenta line) and Long's model prediction (cyan line) of $^3$He/$^4$He fluxes ratio comparing to the data from May 2011 to April 2021 at rigidities = 2.032 GV, 2.531 GV, 3.825 GV and 20.28 GV.}
    \label{fig:He3He4}
\end{figure*}

In Fig. \ref{fig:DHe4} \ref{fig:He3He4}, we show the fluxes ratio of D to $^4$He  and $^3$He to $^4$He at rigidities = 2.032 GV, 2.531 GV, 3.825 GV and 20.28 GV, respectively. The Zhu's model predictions are marked with magenta lines and the Long's model results are marked with cyan lines.  Please note that the flux ratios are not used in the fitting. The model predictions  show that there is nearly no any time-dependent in the  fluxes ratio of D to $^4$He, except at the very low rigidities. While there is a clear time dependence below 3GV for the fluxes ratio of $^3$He to $^4$He.  It appears that the model cannot reproduce the  $^3$He /$^4$He  ratio at low rigidities; however, the difference between the model and the data is no more than 7\%.


According to the Eq. \ref{force_filed}, the flux ratio of particles $a$ and $b$ can be expressed as:
\begin{equation}\label{eq:ratio}
\begin{aligned}
\frac{J_a^{TOA}(R^{TOA})}{J_{b}^{TOA}(R^{TOA})} =& \frac{J_a^{TOA}(E_a^{TOA}) (\frac{Z\beta}{A})_a}{J_{b}^{TOA}(E_{b}^{TOA}) (\frac{Z\beta}{A})_{b}}\\
 =& \frac{(\frac{Z\beta}{A})_a}{(\frac{Z\beta}{A})_{b}}  \frac{J_a^{LIS}(E_a^{LIS})}{J_{b}^{LIS}(E_{b}^{LIS})} \left ( \frac{R_{b}^{LIS}}{R_a^{LIS}} \right ) ^2.
 \end{aligned}
\end{equation}
Here, $E^{TOA} = E^{LIS } - \frac{Ze}{A}\phi(R)$, and $E = \sqrt{R^2(\frac{Ze}{A})^2+m_0^2}-m_0$. 
As the particle have different LIS and Z/A value,  the long-term behavior of the fluxes ratio mainly arises from the second and third term of Eq. \ref{eq:ratio} with the same time series of $\phi(R)$. Because the $^3$He and $^4$He have different Z/A and LIS, the time dependency in the fluxes ratio will be more obvious.
D and $^4$He possess distinct Local Interstellar Spectra (LIS) shapes, attributed to their varied origins \citep{Cooke_2018,Yuan_2024}. Consequently, even though they share the same Z/A, the ratio of D to 
$^4$He could exhibit time dependency in the lower rigidity ranges. This discrepancy highlights the significant influence of LIS on particle fluxes. Similarly, helium exhibits a different LIS compared to carbon and oxygen, whereas carbon and oxygen display remarkably similar LIS \citep{Zhu:2018jbk}. This disparity will result in the He/O ratio showing time dependency at low rigidities, while the C/O ratio remains time-independent. However, the long-term behavior will  beyond the detection capabilities  of AMS-02 \citep{PhysRevLett.134.051001}.

\begin{figure*}
    \centering
    \includegraphics[scale=0.6]{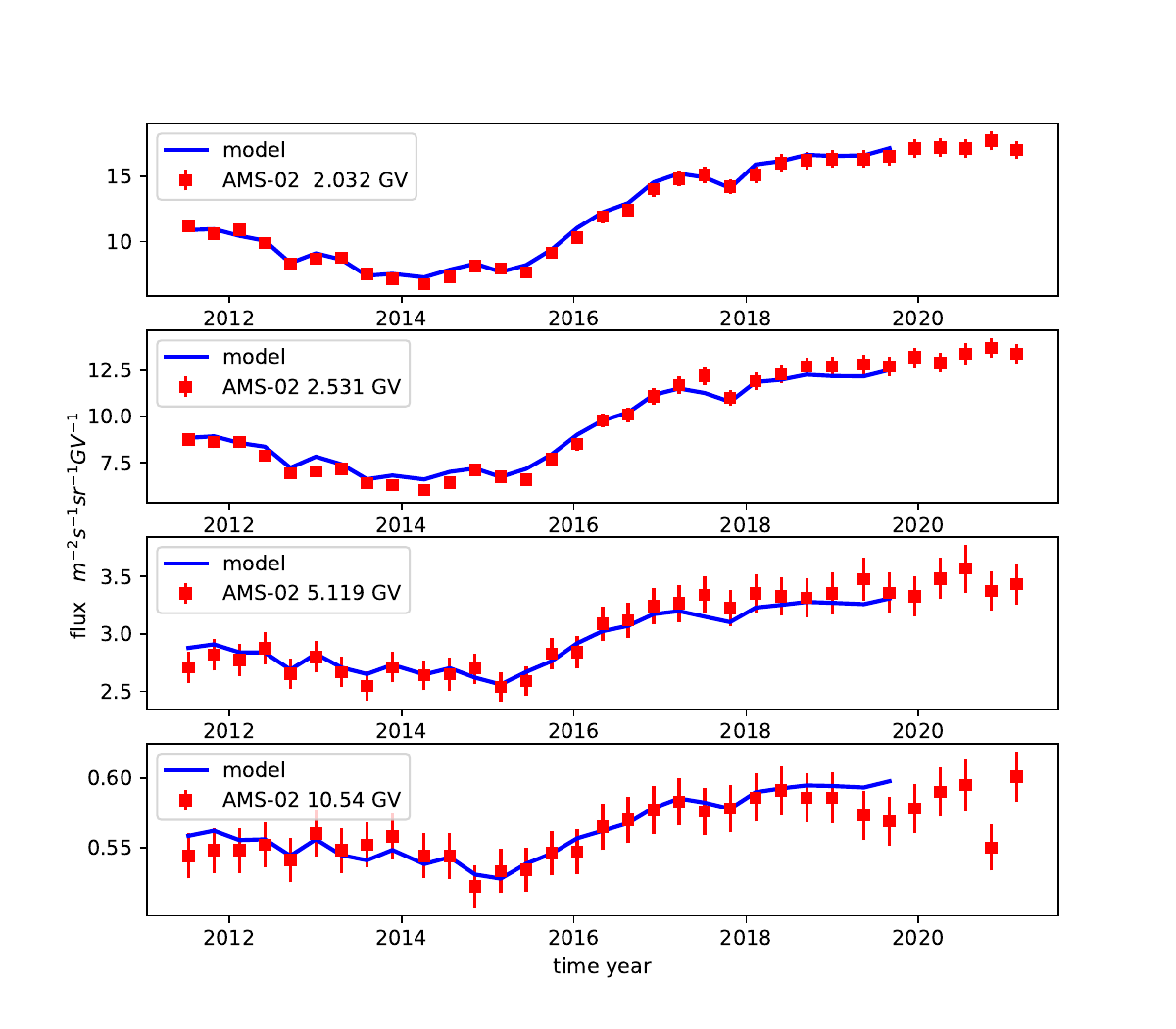}
    \caption{The monthly averaged prediction of D obtained from  daily fluxes (blue lines) of Zhu's model to the data (red points) from May 2011 to April 2021 at rigidities = 2.032 GV, 2.531 GV, 5.119 GV and 10.54 GV.}
    \label{fig:dailyD}
\end{figure*}

\begin{figure*}
    \centering
    \includegraphics[scale=0.6]{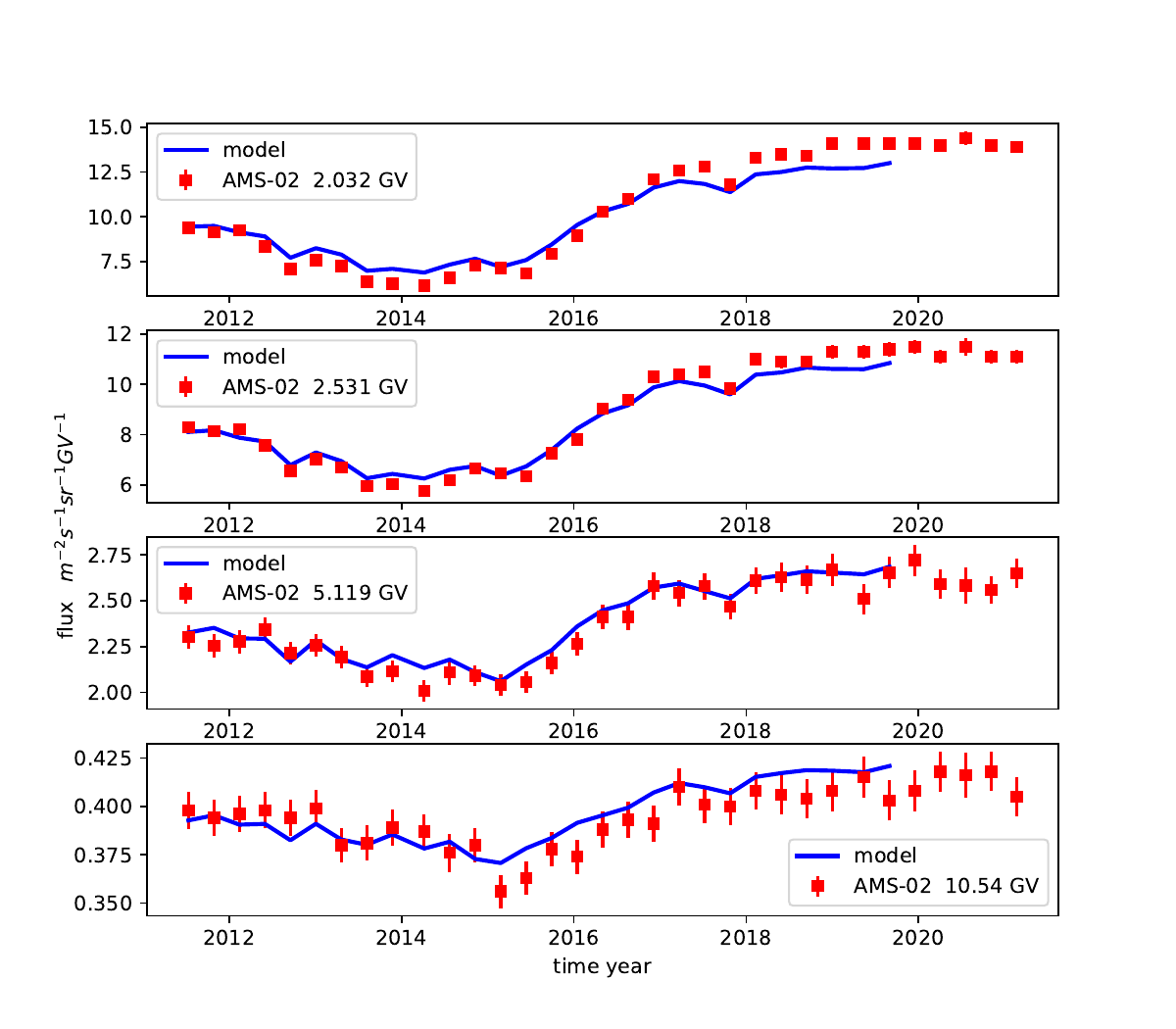}
    \caption{The monthly averaged prediction of $^3$He obtained from  daily fluxes (blue lines)  of Zhu's model  to the data (red points) from May 2011 to April 2021 at rigidities = 2.032 GV, 2.531 GV, 5.119 GV and 10.54 GV.}
    \label{fig:dailyHe3}
\end{figure*}

\begin{figure*}
    \centering
    \includegraphics[scale=0.6]{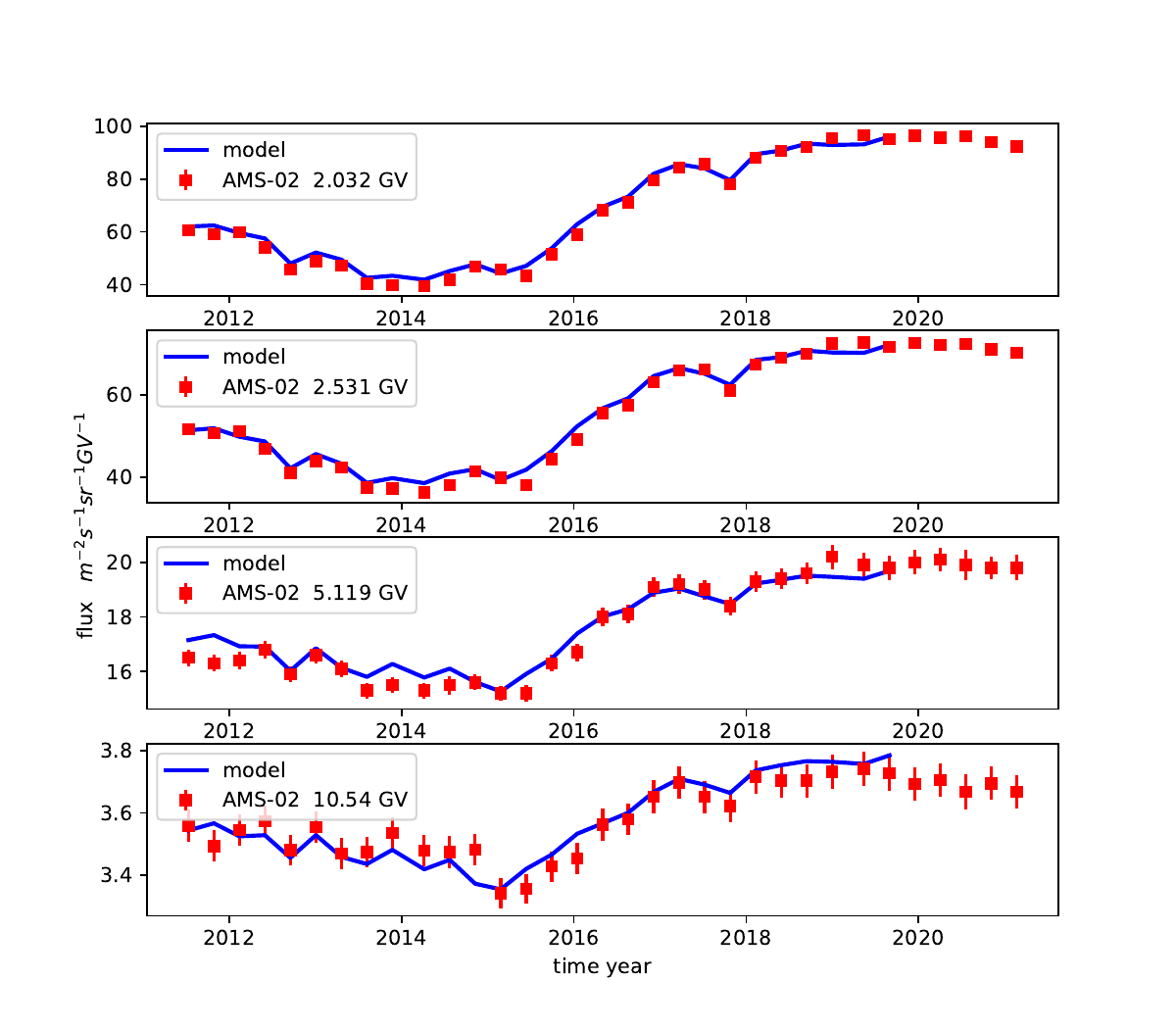}
    \caption{The monthly averaged prediction of $^4$He obtained from Zhu's model  daily fluxes (blue lines)  to the data (red points) from May 2011 to April 2021 at rigidities = 2.032 GV, 2.531 GV, 5.119 GV and 10.54 GV.}
    \label{fig:dailyHe4}
\end{figure*}

\begin{figure}[htbp]
  \centering 
  \begin{minipage}{0.48\textwidth}
    \centering
    \includegraphics[width=\linewidth]{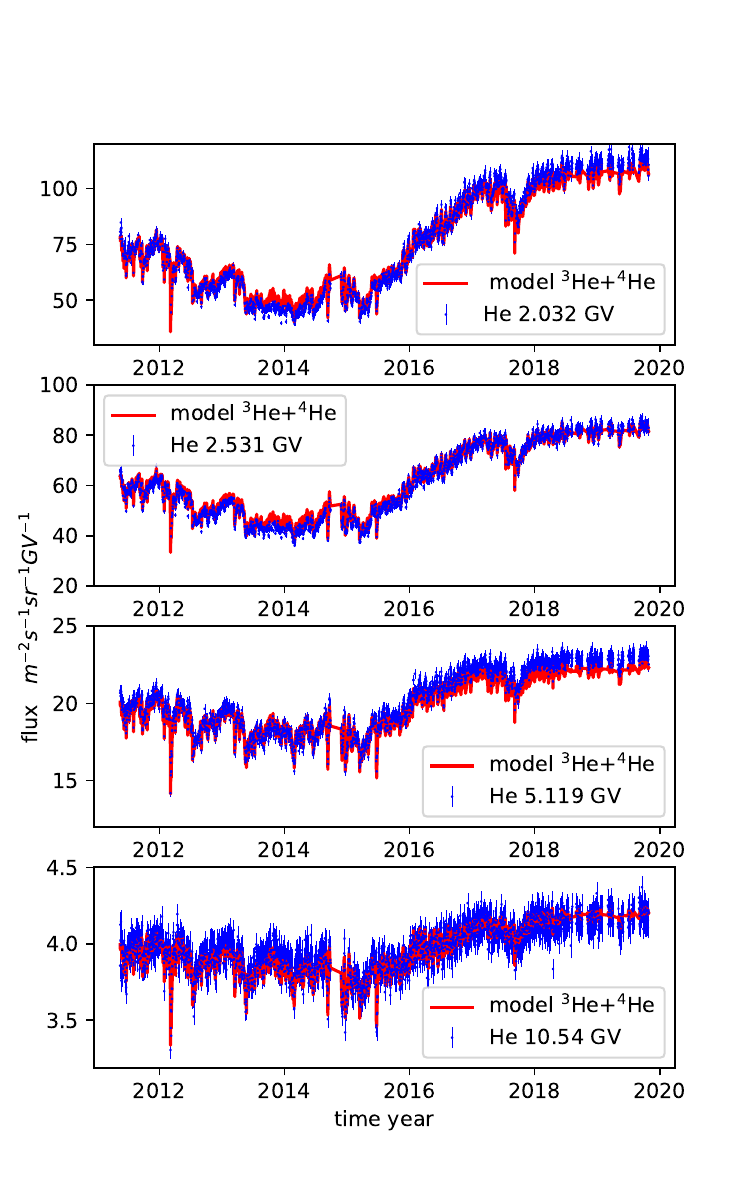}
    \caption{Zhu's model prediction of $^3$He $+$ $^4$He daily fluxes (blue lines)  to the data from 2011 to 2019 \citep{PhysRevLett.128.231102} at rigidities = 2.032 GV, 2.531 GV, 5.119 GV and 10.54 GV.}
    \label{fig:1}
  \end{minipage}
  \hfill
  \begin{minipage}{0.48\textwidth}
    \centering
    \includegraphics[width=\linewidth]{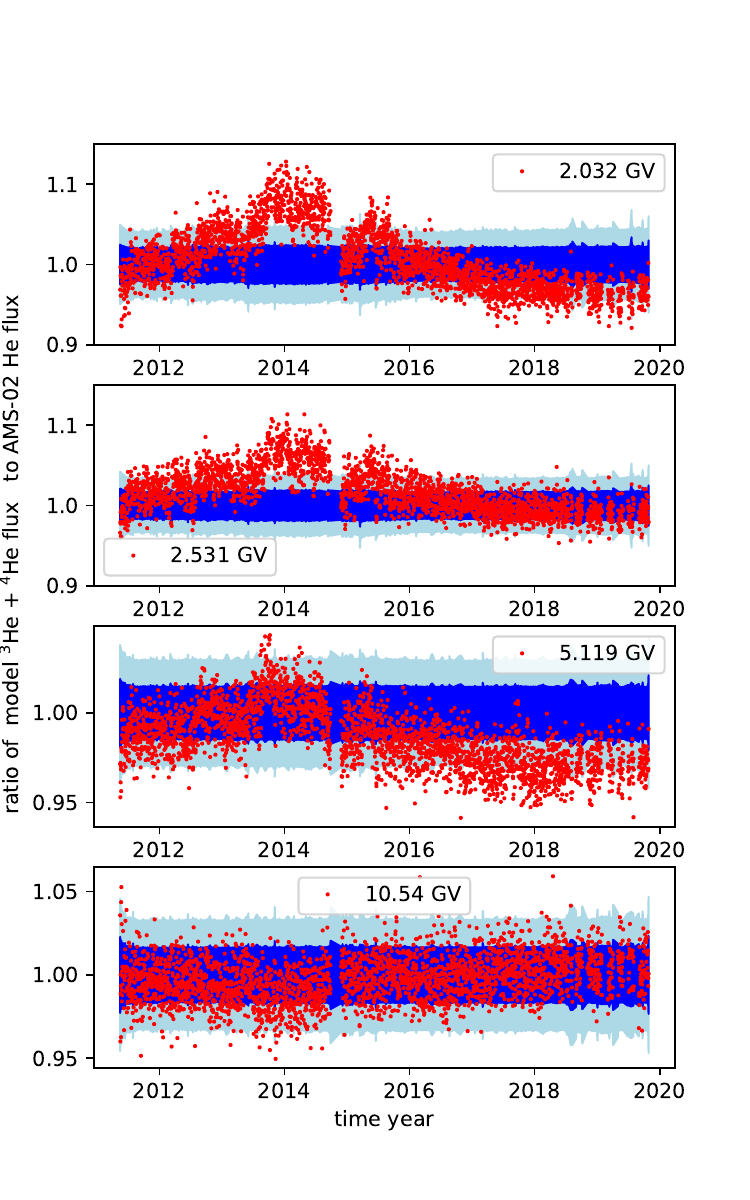}
    \caption{ratio of Zhu's model prediction of $^3$He $+$ $^4$He daily fluxes  to the data (red points)  from 2011 to 2020 \citep{PhysRevLett.128.231102} at rigidities = 2.032 GV, 2.531 GV, 5.119 GV and 10.54 GV. The blue and light blue bands denote the 1 and 2 $\sigma$ CI of measurements.}
    \label{fig:2}
  \end{minipage}
\end{figure}

In our previous work \citep{Zhu_2025}, we obtained the daily solar modulation parameters for p and He. Consequently, we can use these parameters to forecast the daily fluxes of  of D, $^3$He and $^4$He. First, we need the LIS of these three particles, as the parameters from  \citep{Zhu_2025} are deduced with LIS. We assume  that the mean fluxes of  D, $^3$He and $^4$He from 2011/5/20 to 2018/3/29 have same solar modulation potential  as the 7 years period fluxes of  p and He \citep{AGUILAR20211}, which is 0.477 GV. Then we can calculate the daily fluxes of them after get the LIS with the solar modulation parameters. \zhu{We shown the monthly averaged prediction of D, $^3$He and $^4$He obtained from   daily fluxes (blue lines) of Zhu's model  to the data (red points) from May 2011 to April 2021 at rigidities = 2.032 GV, 2.531 GV, 5.119 GV and 10.54 GV in Fig. \ref{fig:dailyD}, \ref{fig:dailyHe3} and \ref{fig:dailyHe4}. The forecasted results are in good agreement with the measurements.} \zhu{In Fig. \ref{fig:1} and \ref{fig:2}, we show the $^3$He fluxes plus $^4$He fluxes comparing to the daily measurement of He from 2011 to 2020 and most of them are consistent with the data within 2\(\sigma\) confidence interval.} \zhu{Our model obtained more fluxes during the max of solar activity around 2014 at the low rigidities, which means our model need to be improved during these period.} For more forecasting results in the range of 2 to 20 GV, please visit our homepage\footnote{ \url{https://github.com/zhucr/daily-fluxes-of-D-He3-He4.git}}.

%


\section{conclusion}

The precise measurement of cosmic ray (CR) spectra is crucial for understanding solar modulation. It also offers a valuable opportunity to enhance our comprehension of CR propagation and to explore new frontiers in astrophysics, and perhaps even uncover new physical phenomena. In this study, we examine the solar modulation of the recently observed time-dependent fluxes of D, 
$^3$He, and $^4$He using data from AMS-02 \citep{PhysRevLett.132.261001} and employing different  modified FFA models. Instead of using a constant solar modulation potential as in the  FFA, they all introduce an rigidity-dependent solar modulation potential $\phi(R)$. Given the current limited understanding of the LIS for these isotopes, we adopt a non-LIS method in our analysis. All the three models can achieve excellent fits to the data using consistent parameters.
Long's model yields the best fitting results with a mean $\chi^2/d.o.f$ value of 0.537. Following that is Zhu's model, which has a mean $\chi^2/d.o.f$  value of 0.771. The Cholis' model provides the highest mean $\chi^2/d.o.f$ value at 1.053.   
      
Combining previous results \citep{PhysRevLett.121.251104, Song_2021,2022PhRvD.106f3006W,PhysRevD.109.083009, zhu2024}, where they fit the proton and Helium fluxes with the same solar modulation parameters, we can achieve excellent fits to the data using consistent parameters across all these isotopes, indicating that these CRs undergo similar propagation processes within the heliosphere. 
These facts prove the prove the assumption in literature that all positively charged CRs undergo the same propagation processes, meaning they all have a universal mean free path.
This assumption also works for the negative particles. In \cite{zhu2025antip}, we predicted the daily fluxes of antiproton, and the subsequent AMS-02 results indicate that the forecasts are in agreement with the measurements within the 2 \(\sigma\) confidence interval \citep{PhysRevLett.134.051002}. In this work, we forecast the daily fluxes of D, $^3$He and $^4$He.
The future time-dependent data from AMS-02 will provide further validation for this assumption.

The time-dependent behaviors of flux ratios at low energies, where the isotopes share the same solar modulation parameters, can be attributed to two main factors: Z/A and LIS . For instance, $^3$He and  $^4$He exhibit different Z/A values and LIS shapes, leading to a time-dependent $^3$He/$^4$He  flux ratio below 3 GV. Similarly, D and $^4$He have distinct LIS shapes, resulting in time-dependent behavior of the D/$^4$He  flux ratio below 4.5 GV, despite their identical Z/A values.

These models are based on a series of assumptions, such as we do not consider the difference in modulation effect from Z/A, which may cause hysteresis between the helium-to-proton flux ratio and the helium flux. This will be further studied in our future work. As the modified FFA models can give very good fitting results, it will be useful for studying the origin and propagation of GCRs in the galaxy.

\begin{acknowledgments}
Thanks to Fan Yi-Zhong, Yuan Qiang and Duan Kai-Kai for very helpful discussions. This work is supported by the National Natural Science Foundation of China  (No. 12203103). Z.C.R is also supported by the Doctoral research start-up funding of Anhui Normal University. We acknowledge the use of  data from the \href{https://ams02.space/publications/}{AMS Publications (https://ams02.space/publications/)}.
\end{acknowledgments}

\section*{Appendix}

\begin{table}
    \centering
    \begin{tabular}{c|c|c|c|c|c}
    \toprule
begin date  & end date &  D  &  $^3$He  & $^4$He   &  D+$^3$He+$^4$He  \\ 
    \midrule
2011-05-20  &  2011-08-30  &  0.139  &  1.100  &  1.589  &  1.315\\ 
2011-08-31  &  2011-12-16  &  0.268  &  1.743  &  1.560  &  1.271\\ 
2011-12-17  &  2012-04-02  &  0.247  &  1.037  &  1.354  &  1.304\\ 
2012-04-03  &  2012-07-19  &  0.123  &  0.625  &  1.231  &  0.914\\ 
2012-07-20  &  2012-11-04  &  0.156  &  1.427  &  1.254  &  1.400\\ 
2012-11-05  &  2013-02-20  &  0.228  &  1.646  &  1.641  &  1.498\\ 
2013-02-21  &  2013-06-08  &  0.134  &  0.379  &  1.151  &  0.900\\ 
2013-06-09  &  2013-09-24  &  0.312  &  1.012  &  2.210  &  1.557\\ 
2013-09-25  &  2014-01-10  &  0.148  &  1.470  &  2.161  &  1.401\\ 
2014-02-07  &  2014-05-25  &  0.221  &  1.361  &  1.599  &  1.389\\ 
2014-05-26  &  2014-09-10  &  0.142  &  0.148  &  0.386  &  0.454\\ 
2014-09-11  &  2014-12-27  &  0.586  &  0.449  &  0.256  &  0.950\\ 
2014-12-28  &  2015-04-14  &  0.146  &  0.343  &  0.052  &  0.393\\ 
2015-04-15  &  2015-07-31  &  0.142  &  0.315  &  0.083  &  0.377\\ 
2015-08-01  &  2015-11-16  &  0.101  &  0.392  &  0.050  &  0.253\\ 
2015-11-17  &  2016-03-03  &  0.069  &  0.311  &  0.092  &  0.202\\ 
2016-03-04  &  2016-06-19  &  0.017  &  0.245  &  0.281  &  0.254\\ 
2016-06-20  &  2016-10-05  &  0.013  &  0.221  &  0.054  &  0.275\\ 
2016-10-06  &  2017-01-21  &  0.061  &  0.213  &  0.185  &  0.412\\ 
2017-01-22  &  2017-05-09  &  0.095  &  1.493  &  0.124  &  0.664\\ 
2017-05-10  &  2017-08-25  &  0.230  &  0.527  &  0.196  &  0.534\\ 
2017-08-26  &  2017-12-11  &  0.092  &  0.250  &  0.136  &  0.316\\ 
2017-12-12  &  2018-03-29  &  0.107  &  0.393  &  0.133  &  0.621\\ 
2018-03-30  &  2018-07-15  &  0.156  &  0.644  &  0.177  &  0.545\\ 
2018-07-16  &  2018-10-31  &  0.172  &  0.746  &  0.120  &  0.403\\ 
2018-11-01  &  2019-02-16  &  0.201  &  0.220  &  0.162  &  0.385\\ 
2019-03-16  &  2019-07-01  &  0.267  &  0.948  &  0.163  &  0.742\\ 
2019-07-02  &  2019-10-17  &  0.105  &  0.743  &  0.120  &  0.603\\ 
2019-10-18  &  2020-02-02  &  0.220  &  0.153  &  0.308  &  0.456\\ 
2020-02-03  &  2020-05-20  &  0.105  &  0.901  &  0.166  &  0.623\\ 
2020-05-21  &  2020-09-05  &  0.269  &  0.585  &  0.269  &  0.811\\ 
2020-09-06  &  2020-12-22  &  0.361  &  0.732  &  0.360  &  1.283\\ 
2020-12-23  &  2021-04-09  &  0.463  &  0.317  &  0.307  &  0.955\\ 

    \bottomrule
    \end{tabular}
    \caption{$\chi^2/d.o.f$ from fitting to D, $^3$He, $^4$He individually and fitting to D, $^3$He and $^4$He simultaneously with with Zhu's model.}
    \label{tab:zhu}
\end{table}

\begin{table}
    \centering
    \begin{tabular}{c|c|c|c|c|c}
    \toprule
begin date  & end date &  D  &  $^3$He  & $^4$He   &  D+$^3$He+$^4$He  \\ 
    \midrule
2011-05-20  &  2011-08-30  &  0.133  &  1.080  &  2.570  &  1.555\\ 
2011-08-31  &  2011-12-16  &  0.274  &  1.687  &  1.665  &  1.290\\ 
2011-12-17  &  2012-04-02  &  0.248  &  0.977  &  1.747  &  1.260\\ 
2012-04-03  &  2012-07-19  &  0.250  &  0.411  &  0.966  &  0.977\\ 
2012-07-20  &  2012-11-04  &  0.324  &  1.068  &  0.974  &  1.487\\ 
2012-11-05  &  2013-02-20  &  0.889  &  1.122  &  0.872  &  1.597\\ 
2013-02-21  &  2013-06-08  &  0.411  &  0.282  &  0.702  &  0.966\\ 
2013-06-09  &  2013-09-24  &  1.244  &  0.731  &  1.986  &  1.993\\ 
2013-09-25  &  2014-01-10  &  0.797  &  1.198  &  2.435  &  2.239\\ 
2014-02-07  &  2014-05-25  &  0.536  &  1.362  &  2.757  &  2.361\\ 
2014-05-26  &  2014-09-10  &  0.570  &  0.099  &  1.163  &  0.964\\ 
2014-09-11  &  2014-12-27  &  0.641  &  0.254  &  0.962  &  1.170\\ 
2014-12-28  &  2015-04-14  &  0.355  &  0.396  &  0.298  &  0.460\\ 
2015-04-15  &  2015-07-31  &  0.304  &  0.277  &  0.394  &  0.462\\ 
2015-08-01  &  2015-11-16  &  0.208  &  0.290  &  0.120  &  0.298\\ 
2015-11-17  &  2016-03-03  &  0.081  &  0.298  &  0.285  &  0.271\\ 
2016-03-04  &  2016-06-19  &  0.023  &  0.567  &  0.368  &  0.304\\ 
2016-06-20  &  2016-10-05  &  0.016  &  0.368  &  0.315  &  0.405\\ 
2016-10-06  &  2017-01-21  &  0.061  &  0.737  &  0.625  &  0.647\\ 
2017-01-22  &  2017-05-09  &  0.084  &  1.497  &  0.472  &  0.751\\ 
2017-05-10  &  2017-08-25  &  0.289  &  0.799  &  0.807  &  0.921\\ 
2017-08-26  &  2017-12-11  &  0.108  &  0.289  &  0.150  &  0.337\\ 
2017-12-12  &  2018-03-29  &  0.107  &  0.541  &  0.240  &  0.719\\ 
2018-03-30  &  2018-07-15  &  0.168  &  0.642  &  0.568  &  0.697\\ 
2018-07-16  &  2018-10-31  &  0.206  &  0.692  &  0.640  &  0.593\\ 
2018-11-01  &  2019-02-16  &  0.214  &  0.205  &  1.041  &  0.692\\ 
2019-03-16  &  2019-07-01  &  0.739  &  1.035  &  0.515  &  0.898\\ 
2019-07-02  &  2019-10-17  &  0.462  &  0.972  &  0.561  &  1.001\\ 
2019-10-18  &  2020-02-02  &  0.376  &  0.566  &  2.459  &  1.288\\ 
2020-02-03  &  2020-05-20  &  0.107  &  0.841  &  1.558  &  1.043\\ 
2020-05-21  &  2020-09-05  &  0.252  &  0.528  &  2.065  &  1.389\\ 
2020-09-06  &  2020-12-22  &  2.913  &  0.752  &  2.257  &  2.172\\ 
2020-12-23  &  2021-04-09  &  0.473  &  0.298  &  2.367  &  1.564\\

    \bottomrule
    \end{tabular}
    \caption{$\chi^2/d.o.f$ from fitting to D, $^3$He, $^4$He individually and fitting to D, $^3$He and $^4$He simultaneously with Cholis' model.}
    \label{tab:Cholis}
\end{table}

\begin{table}
    \centering
    \begin{tabular}{c|c|c|c|c|c}
    \toprule
begin date  & end date &  D  &  $^3$He  & $^4$He   &  D+$^3$He+$^4$He  \\ 
    \midrule
2011-05-20  &  2011-08-30  &  0.158  &  0.923  &  1.436  &  1.224\\ 
2011-08-31  &  2011-12-16  &  0.261  &  1.171  &  1.518  &  1.239\\ 
2011-12-17  &  2012-04-02  &  0.225  &  0.817  &  1.085  &  1.076\\ 
2012-04-03  &  2012-07-19  &  0.204  &  0.337  &  0.075  &  0.534\\ 
2012-07-20  &  2012-11-04  &  0.250  &  0.917  &  0.052  &  0.909\\ 
2012-11-05  &  2013-02-20  &  0.441  &  0.353  &  0.199  &  0.810\\ 
2013-02-21  &  2013-06-08  &  0.231  &  0.569  &  0.173  &  0.541\\ 
2013-06-09  &  2013-09-24  &  0.485  &  0.254  &  0.106  &  0.525\\ 
2013-09-25  &  2014-01-10  &  0.229  &  0.285  &  0.087  &  0.408\\ 
2014-02-07  &  2014-05-25  &  0.332  &  0.207  &  0.105  &  0.548\\ 
2014-05-26  &  2014-09-10  &  0.295  &  0.433  &  0.141  &  0.357\\ 
2014-09-11  &  2014-12-27  &  0.529  &  0.228  &  0.292  &  0.870\\ 
2014-12-28  &  2015-04-14  &  0.233  &  0.455  &  0.127  &  0.335\\ 
2015-04-15  &  2015-07-31  &  0.188  &  0.793  &  0.125  &  0.373\\ 
2015-08-01  &  2015-11-16  &  0.155  &  0.588  &  0.241  &  0.351\\ 
2015-11-17  &  2016-03-03  &  0.074  &  0.289  &  0.080  &  0.182\\ 
2016-03-04  &  2016-06-19  &  0.019  &  0.191  &  0.112  &  0.205\\ 
2016-06-20  &  2016-10-05  &  0.022  &  0.085  &  0.046  &  0.247\\ 
2016-10-06  &  2017-01-21  &  0.079  &  0.386  &  0.081  &  0.340\\ 
2017-01-22  &  2017-05-09  &  0.098  &  1.382  &  0.062  &  0.530\\ 
2017-05-10  &  2017-08-25  &  0.218  &  0.083  &  0.078  &  0.376\\ 
2017-08-26  &  2017-12-11  &  0.094  &  0.090  &  0.045  &  0.241\\ 
2017-12-12  &  2018-03-29  &  0.109  &  0.214  &  0.043  &  0.421\\ 
2018-03-30  &  2018-07-15  &  0.140  &  0.221  &  0.063  &  0.359\\ 
2018-07-16  &  2018-10-31  &  0.185  &  0.353  &  0.082  &  0.281\\ 
2018-11-01  &  2019-02-16  &  0.198  &  0.189  &  0.202  &  0.292\\ 
2019-03-16  &  2019-07-01  &  0.230  &  1.226  &  0.122  &  0.576\\ 
2019-07-02  &  2019-10-17  &  0.214  &  0.206  &  0.124  &  0.424\\ 
2019-10-18  &  2020-02-02  &  0.266  &  0.231  &  0.167  &  0.291\\ 
2020-02-03  &  2020-05-20  &  0.095  &  0.590  &  0.129  &  0.457\\ 
2020-05-21  &  2020-09-05  &  0.302  &  0.323  &  0.117  &  0.651\\ 
2020-09-06  &  2020-12-22  &  0.721  &  0.762  &  0.085  &  0.918\\ 
2020-12-23  &  2021-04-09  &  0.438  &  0.142  &  0.174  &  0.828\\ 

    \bottomrule
    \end{tabular}
    \caption{$\chi^2/d.o.f$ from fitting to D, $^3$He, $^4$He individually and fitting to D, $^3$He and $^4$He simultaneously with Long's model.}
    \label{tab:Long}
\end{table}

\begin{table}
    \centering
    \begin{tabular}{c|c|c|c|c|c}
    \toprule
begin date  & end date &  D  &  $^3$He  & $^4$He   &  D+$^3$He+$^4$He  \\ 
    \midrule
2011-05-20  &  2011-08-30  &  0.311  &  1.611  &  3.223  &  1.819\\ 
2011-08-31  &  2011-12-16  &  0.258  &  2.447  &  1.609  &  1.438\\ 
2011-12-17  &  2012-04-02  &  0.262  &  2.259  &  2.627  &  1.789\\ 
2012-04-03  &  2012-07-19  &  0.465  &  5.439  &  6.016  &  3.924\\ 
2012-07-20  &  2012-11-04  &  0.628  &  9.034  &  8.132  &  5.855\\ 
2012-11-05  &  2013-02-20  &  2.049  &  9.064  &  10.426  &  7.253\\ 
2013-02-21  &  2013-06-08  &  0.732  &  4.468  &  6.548  &  4.087\\ 
2013-06-09  &  2013-09-24  &  2.139  &  9.110  &  15.245  &  8.915\\ 
2013-09-25  &  2014-01-10  &  3.866  &  14.464  &  22.243  &  13.382\\ 
2014-02-07  &  2014-05-25  &  2.858  &  10.931  &  18.440  &  10.866\\ 
2014-05-26  &  2014-09-10  &  1.700  &  4.410  &  10.228  &  5.358\\ 
2014-09-11  &  2014-12-27  &  0.601  &  3.400  &  4.519  &  2.967\\ 
2014-12-28  &  2015-04-14  &  0.503  &  0.849  &  0.791  &  0.799\\ 
2015-04-15  &  2015-07-31  &  0.654  &  1.664  &  2.565  &  1.621\\ 
2015-08-01  &  2015-11-16  &  0.404  &  0.923  &  0.905  &  0.762\\ 
2015-11-17  &  2016-03-03  &  0.082  &  0.331  &  0.324  &  0.265\\ 
2016-03-04  &  2016-06-19  &  0.099  &  0.536  &  0.376  &  0.329\\ 
2016-06-20  &  2016-10-05  &  0.101  &  0.884  &  0.315  &  0.433\\ 
2016-10-06  &  2017-01-21  &  0.132  &  1.001  &  0.594  &  0.646\\ 
2017-01-22  &  2017-05-09  &  0.109  &  1.766  &  0.503  &  0.787\\ 
2017-05-10  &  2017-08-25  &  0.290  &  2.132  &  1.625  &  1.414\\ 
2017-08-26  &  2017-12-11  &  0.099  &  0.662  &  0.417  &  0.488\\ 
2017-12-12  &  2018-03-29  &  0.161  &  1.099  &  0.402  &  0.798\\ 
2018-03-30  &  2018-07-15  &  0.173  &  1.453  &  1.053  &  0.952\\ 
2018-07-16  &  2018-10-31  &  0.193  &  1.382  &  1.094  &  0.875\\ 
2018-11-01  &  2019-02-16  &  0.200  &  1.118  &  1.142  &  0.827\\ 
2019-03-16  &  2019-07-01  &  0.687  &  1.944  &  0.786  &  1.125\\ 
2019-07-02  &  2019-10-17  &  0.732  &  2.169  &  1.119  &  1.469\\ 
2019-10-18  &  2020-02-02  &  0.457  &  1.491  &  3.595  &  1.884\\ 
2020-02-03  &  2020-05-20  &  0.104  &  2.101  &  2.084  &  1.418\\ 
2020-05-21  &  2020-09-05  &  0.336  &  1.905  &  3.345  &  1.922\\ 
2020-09-06  &  2020-12-22  &  4.148  &  1.973  &  3.304  &  3.111\\ 
2020-12-23  &  2021-04-09  &  0.502  &  1.277  &  3.404  &  1.970\\

    \bottomrule
    \end{tabular}
    \caption{$\chi^2/d.o.f$ from fitting to D, $^3$He, $^4$He individually and fitting to D, $^3$He and $^4$He simultaneously with FFA.}
    \label{tab:FFA}
\end{table}

\clearpage

\bibliographystyle{aasjournal}
\bibliography{refs}{}

\end{document}